\documentclass[letterpaper, 10 pt, journal, twoside]{ieeetran}

\IEEEoverridecommandlockouts                              

\usepackage{stfloats}
\usepackage[utf8]{inputenc}
\usepackage{graphicx} 
\usepackage{color}
\usepackage{comment}
\usepackage{amssymb} 
\usepackage{amsmath}
\usepackage{multirow}
\usepackage{float}
\usepackage{bm}
\usepackage{cancel}
\usepackage{subfigure}
\usepackage[dvipsnames]{xcolor}
\usepackage{cite}
\usepackage[top=2.01cm, bottom=1.52cm, left=1.69cm, right=1.69cm]{geometry}
\usepackage[absolute,overlay]{textpos} 

\usepackage{algorithm}
\usepackage[noend]{algpseudocode}

\title{\LARGE \bf
SOPHIE: SOft and flexible aerial vehicle for PHysical Interaction with the Environment
}

\author{F. Ruiz, B.C. Arrue* and A. Ollero
\thanks{F. Ruiz (frvincueria@us.es), B.C. Arrue*
(barrue@us.es) and A. Ollero
(aollero@us.es) are with the GRVC Robotics Lab of Seville, Spain.
}%
}

\begin{document}

\maketitle

\begin{textblock*}{20cm}(0.5cm,27cm) 
  \scriptsize © 2022 IEEE.  Personal use of this material is permitted.  Permission from IEEE must be obtained for all other uses, in any current or future media, including reprinting/republishing this material for advertising or promotional purposes, creating new collective works, for resale or redistribution to servers or lists, or reuse of any copyrighted component of this work in other works.
\end{textblock*}

\begin{abstract}

 This paper presents the first design of a soft, 3D-printed in flexible filament, lightweight UAV, capable of performing full-body perching using soft tendons, specifically landing and stabilizing on pipelines and irregular surfaces without the need for an auxiliary system. The flexibility of the UAV can be controlled during the additive manufacturing process by adjusting the infill rate $\rho_{TPU}$ distribution. However, the increase in flexibility implies difficulties in controlling the UAV, as well as structural, aerodynamic, and aeroelastic effects. This article provides insight into the dynamics of the system and validates the flyability of the vehicle for densities as low as 6$\%$. Within this range, quasi-static arm deformations can be considered, thus the autopilot is fed back through a static arm deflection model. At lower densities, strong non-linear elastic dynamics appear, which translates to complex modeling, and it is suggested to switch to data-based approaches.

\end{abstract}

\section{INTRODUCTION}

Unmanned Aerial Vehicles (UAVs) will have an important impact on society in the short-medium term \cite{Rao2016}. They have the potential to change the way in which various tasks are accomplished and dramatically alter several industries. It has been clearly demonstrated that this technology offers a cost-effective solution to several operations such as surveillance, monitoring, and inspection  \cite{Bernard2011AutonomousTA,OlleroManipulation}. Its high level of autonomy, combined with its ability to access difficult-to-reach areas, including targets at altitude, are some of the reasons. They have the possibility of physically interacting with the environment, for example through robotic arms \cite{JimenezCano2013ControlOA, Manipulators}.

It has been recently highlighted that these systems can be more efficient if they work in cooperation with human beings \cite{ZHAO2020,Zheng2021EvolutionaryHC}. However, traditional UAVs have rigid underlying structures that might possess an inherent risk when in contact with humans. Soft robotics is dedicated to the design and construction of robots with physically flexible bodies \cite{Rus2015DesignFA,Trivedi2008SoftRB}. Nevertheless, in aerial robotics, this trend has only been applied to certain parts of the platform. This is the case of deformable propellers \cite{deformablepropellers} and micro quadcopters with arms that absorb energy in case of impact \cite{Floreano_origami, MINICORE}. 

UAVs can also allow access to zones which are unreachable for humans and traditional UAVs, by modifying their own geometry during flight. This technique is known as morphing \cite{Riviere2018AgileRF} and can also be used to improve the aerodynamic properties and flight mechanics of the UAV \cite{Ajanic2020BioInspiredSW}. The most widespread concepts for these changes in geometry are reversible plasticity, which can be achieved by temperature control \cite{ReversiblePlasticity}, and origami techniques \cite{iros-2016, CargoDrone, origamiQuadrotor}.

\begin{figure}
  \begin{subfigure}[Natural state]{
    \begin{minipage}[t]{0.48\linewidth}
    \centering
    \includegraphics[width=\textwidth,scale=1]{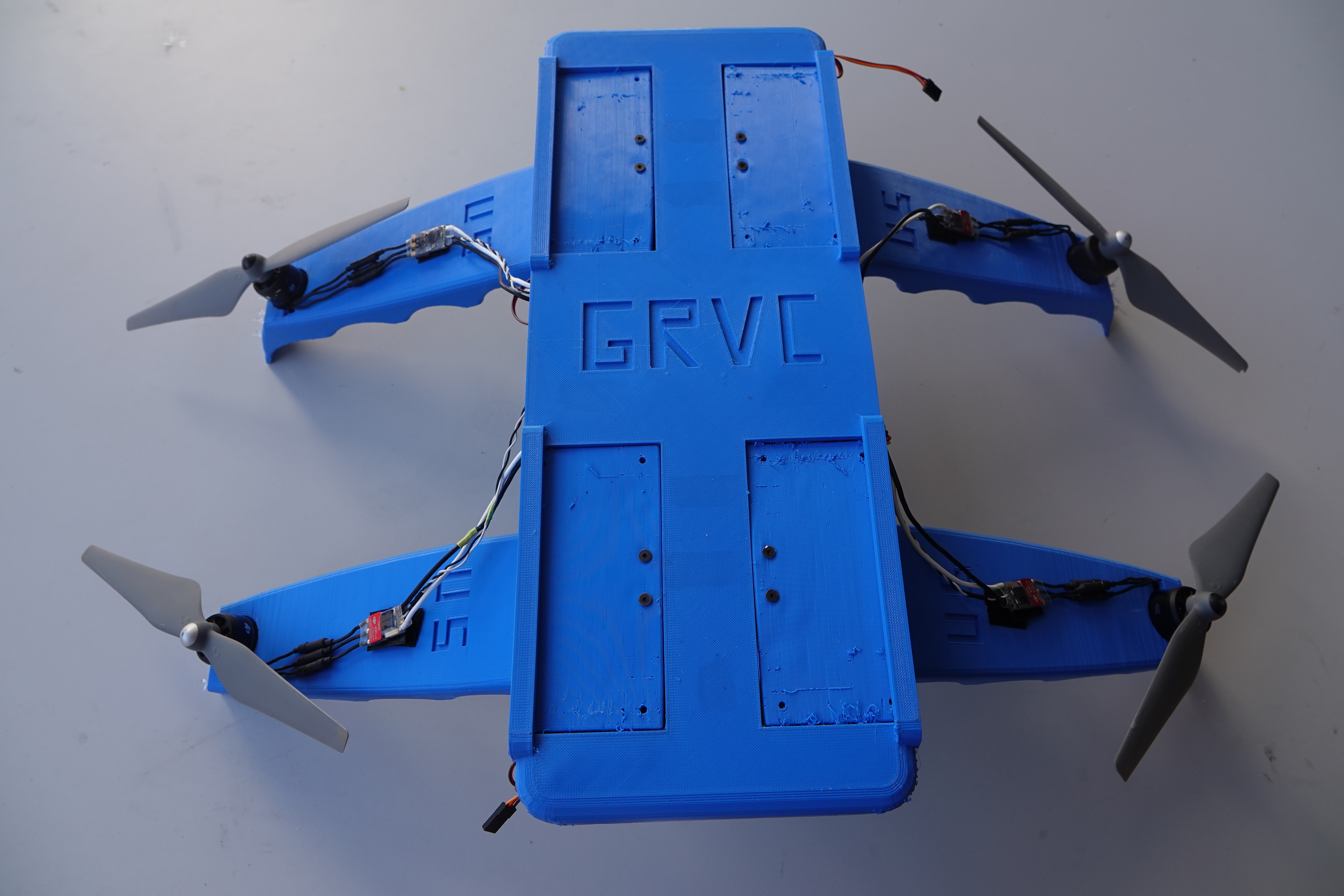}
    \end{minipage}%
    }%
  \end{subfigure}
  \begin{subfigure}[Flexed state]{
    \begin{minipage}[t]{0.48\linewidth}
    \centering
    \includegraphics[width=\textwidth,scale=1]{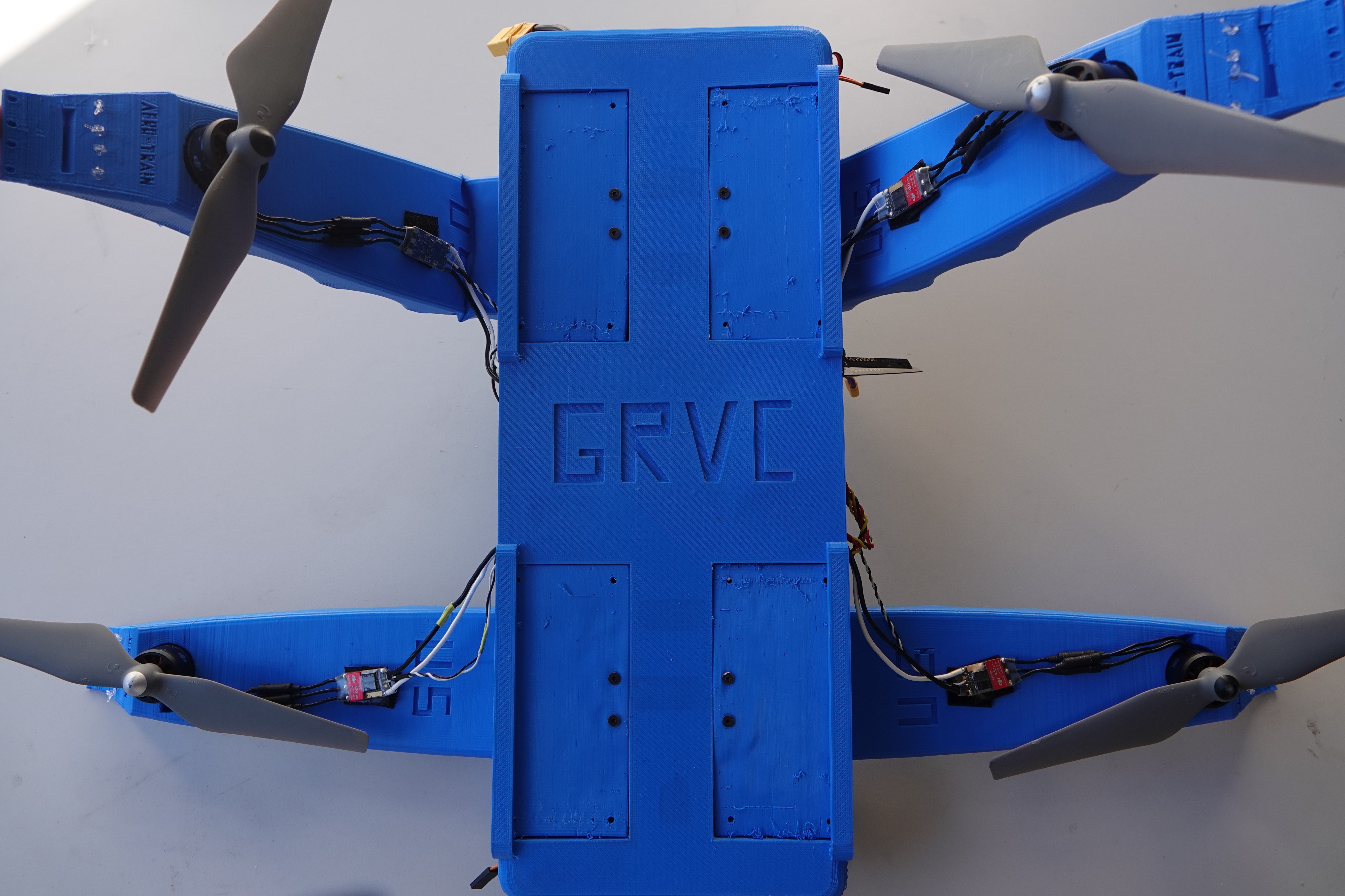}
    \end{minipage}%
    }%
  \end{subfigure}
  \begin{subfigure}[Adapted to the industrial environment]{
    \begin{minipage}[t]{0.96\linewidth}
    \centering
    \includegraphics[width=0.4\textwidth]{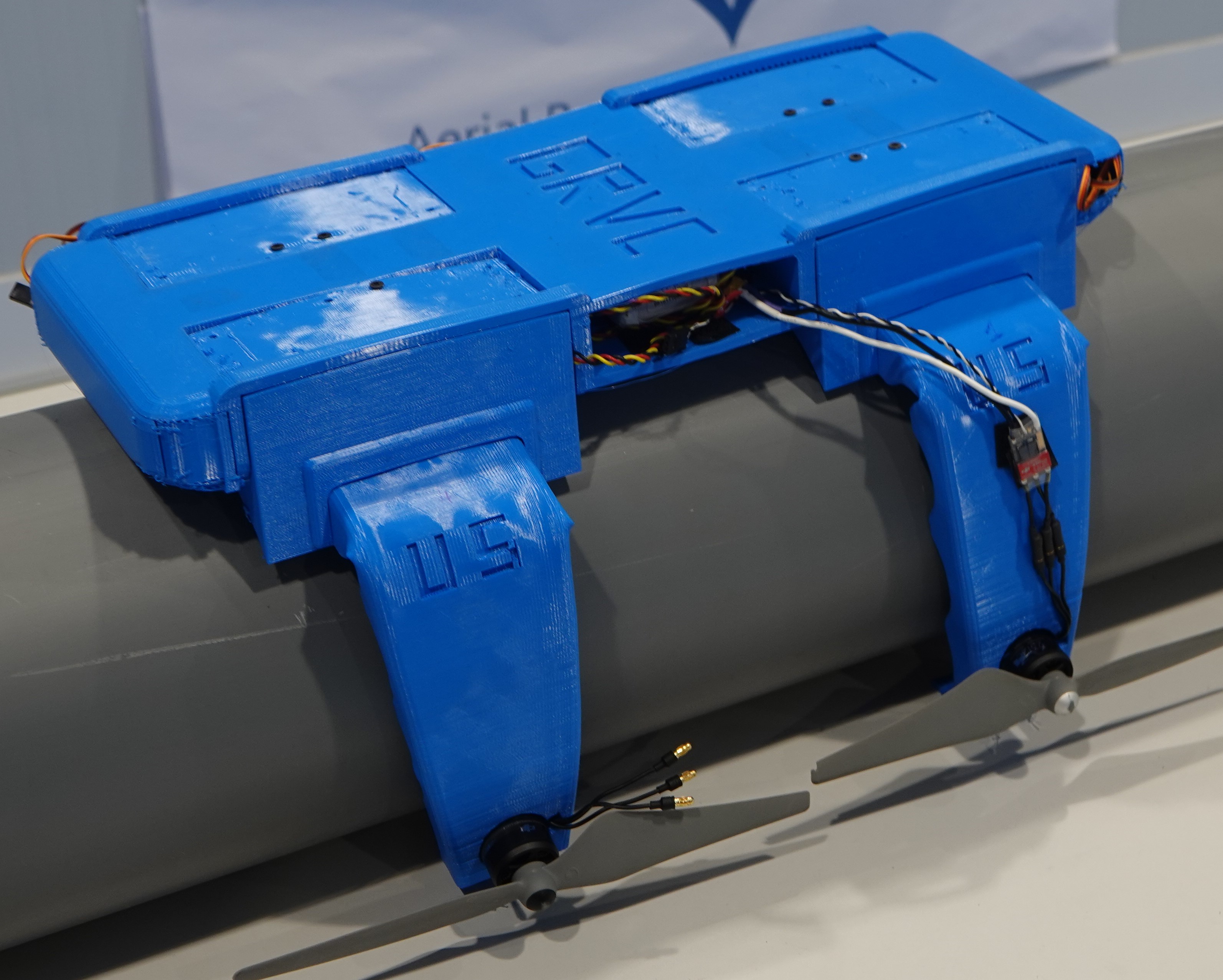}
    \end{minipage}%
    }%
  \end{subfigure}
 \caption{Soft aerial robot prototype entirely 3D-printed in flexible TPU 70A with an internal density $\rho_{TPU}=6.9\%$.}
 \label{f:UAV}
\end{figure}

The use of flexible components is not limited to the structure. For instance, robotic arms can employ soft end-effectors for contact manipulation  \cite{Kovac2019soft,Xiang2019SoftsmartRE}. Soft actuators can provide a greater force-weight ratio than traditional servos, for example TCP tensors \cite{TammTCP}. Moreover, soft systems can be used to perform perching on complex structures or pipelines \cite{GRubiales2021}. 

Soft components are typically 3D-printed using novel materials such as Filaflex or Thermoplastic polyurethane (TPU), which are flexible and elastic. Other soft structures \cite{Mazzolai2012} are based on highly flexible silicones, like Ecoflex. 

In the design of soft aerial robotic systems, nature is the main source of inspiration \cite{Kova2016LearningFN, KovacAdaptive}. Animals are mainly composed of soft components used to move efficiently and safely in different environments. However, the manual design and, especially, the control of these soft robots is an arduous task. The very fact that they are flexible increases significantly the mathematical complexities in the modeling and control when comparing  to  rigid robots \cite{ControlTentacles}. To solve and understand these strong non-linear behaviors, different artificial intelligence methodologies have been proposed to date \cite{Kim2021}. 

The difficulty aforementioned is the reason why completely soft UAVs have not been developed to date. To the authors best knowledge, this article (which is a continuation of the soft propelled arm developed in \cite{softArm}) presents a first of its kind, 100 $\%$ soft structure multirotor, 3D-printed entirely in flexible material (TPU), with a significant weight reduction compared to a conventional multirotor. The flexibility of the UAV can be controlled during the additive manufacturing process by adjusting the internal density of the material $\rho_{TPU}$. 

\begin{figure}
  \begin{subfigure}[Flight phase]{
    \begin{minipage}[t]{0.48\linewidth}
    \centering
    \includegraphics[width=\textwidth,scale=1]{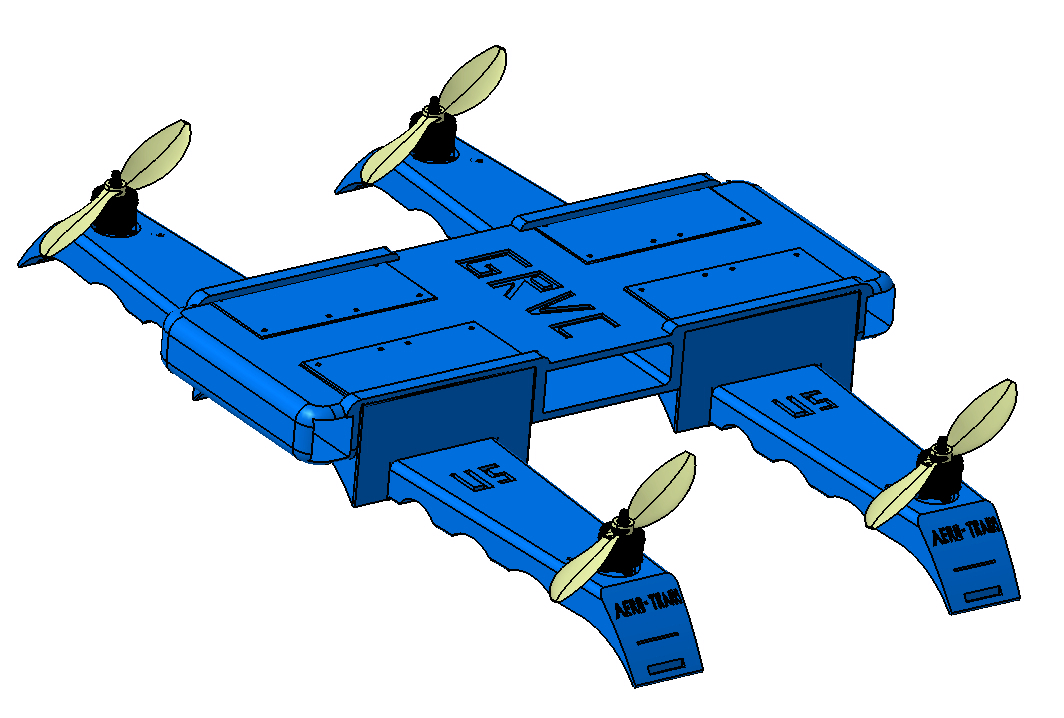}
    \end{minipage}%
    }%
  \end{subfigure}
  \begin{subfigure}[Landing]{
    \begin{minipage}[t]{0.48\linewidth}
    \centering
    \includegraphics[width=\textwidth,scale=1]{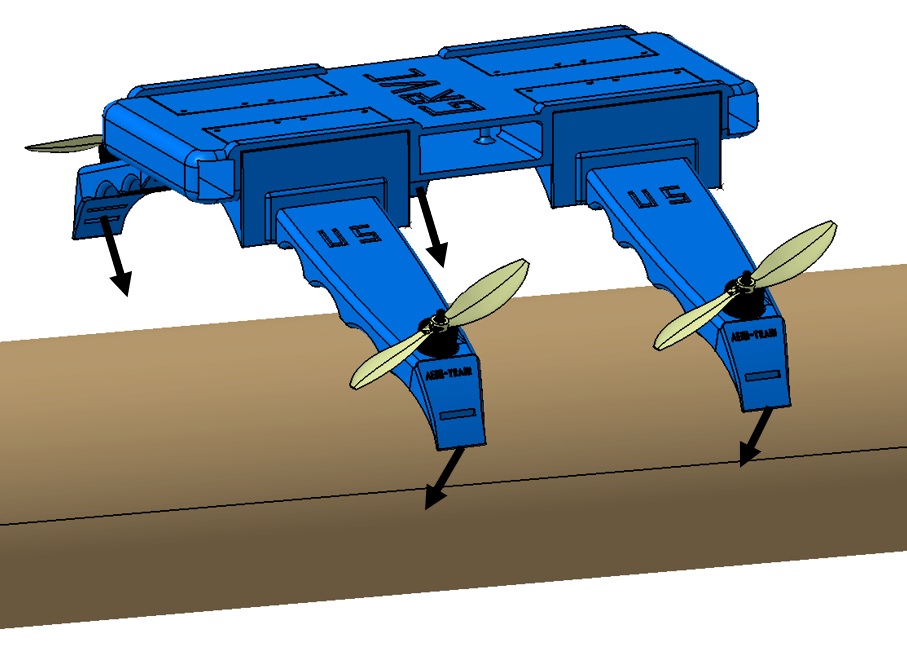}
    \end{minipage}%
    }
  \end{subfigure}
  \begin{subfigure}[Inspection]{
    \begin{minipage}[t]{0.47\linewidth}
    \centering
    \includegraphics[width=\textwidth,scale=1]{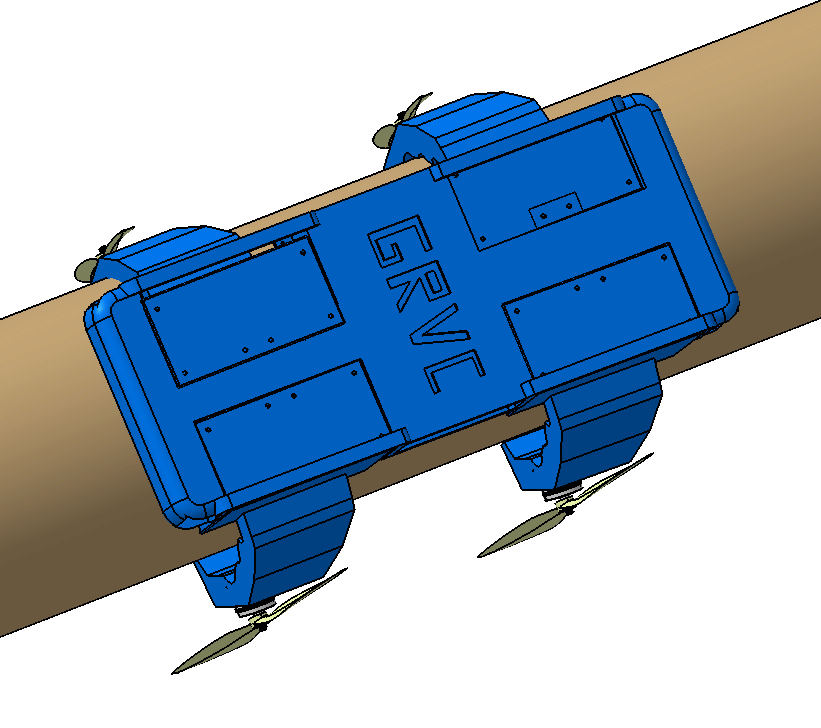}
    \end{minipage}%
    }
  \end{subfigure}
  \begin{subfigure}[Take-off]{
    \begin{minipage}[t]{0.47\linewidth}
    \centering
    \includegraphics[width=\textwidth,scale=1]{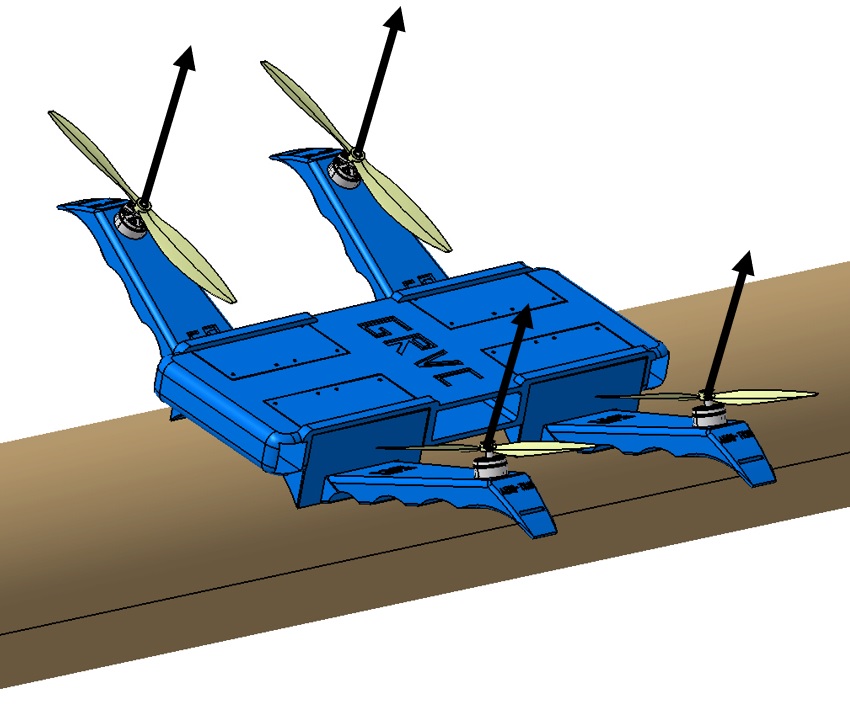}
    \end{minipage}%
    }
  \end{subfigure}
  \label{operation}
  \caption{Concept of operation: approximation and landing on the
pipe, inspection tasks and take-off.}
\vspace{-5mm}
\end{figure}


In addition to the inherent increase in safety in the event of a collision, this UAV concept might have applications such as grasping and perching. In this work, the aerial vehicle has been equipped with a landing system for pipelines based on soft tendons to attach to the surface. This application has emerged as a continuation of the work of the authors within the framework of the European project HYFLIERS.

Using manipulators to attach the UAV to a surface is still challenging nowadays. Authors in \cite{Mattar2018} developed a drone which stuck to magnetic surfaces in refineries to carry out inspection. Subsequently, solutions based on soft robotics \cite{ScienceRoderick} have been chosen, including previous works developed by the authors of the present design \cite{GRubiales2021, PRamon2019}. These bio-inspired approaches are naturally efficient. However, until now they have been designed as add-ons, auxiliary systems that are incorporated into a traditional rigid UAV, adding an enormous extra weight. In this work, the aerial robot itself has the ability, independently, to attach to the pipe thanks to its flexible nature. 

The paper is structured as follows: Section II
describes and justifies the UAV concept presented. Section III details the
UAV design and mechanical considerations. Section IV details the dynamics of the system and the control architecture. Section V validates the flyability of the vehicle and the ability to land on pipes.
Conclusions and new lines of research are proposed in Section VI.

\section{System description}

\subsection{Problem statement} 


The use of soft materials is very advantageous for any aerial robot. One question that arises in this work is to what extent the effects of said flexibility are negligible for the autopilot. Then, once inside said zone of important effects, to what extent can they be resolved with models and control techniques. Finally, the maximum level of flexibility requires machine learning techniques.

\begin{figure}[!tbp]
  \begin{subfigure}[Lower Surface]{
    \begin{minipage}[t]{0.48\linewidth}
    \centering
    \includegraphics[width=\textwidth,scale=1]{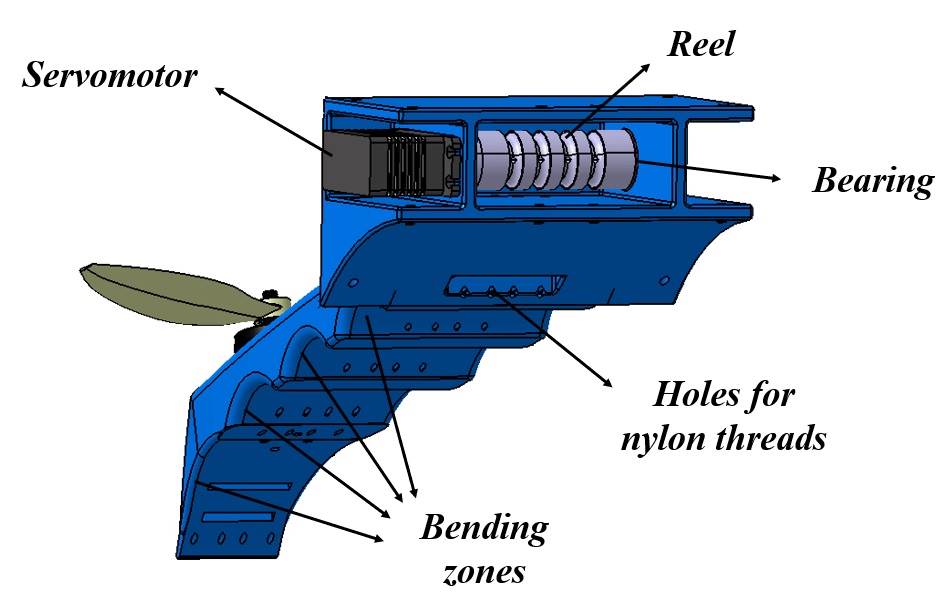}
    \end{minipage}%
    }%
  \end{subfigure}
    \begin{subfigure}[Upper Surface]{
    \begin{minipage}[t]{0.48\linewidth}
    \centering
    \includegraphics[width=\textwidth,scale=1]{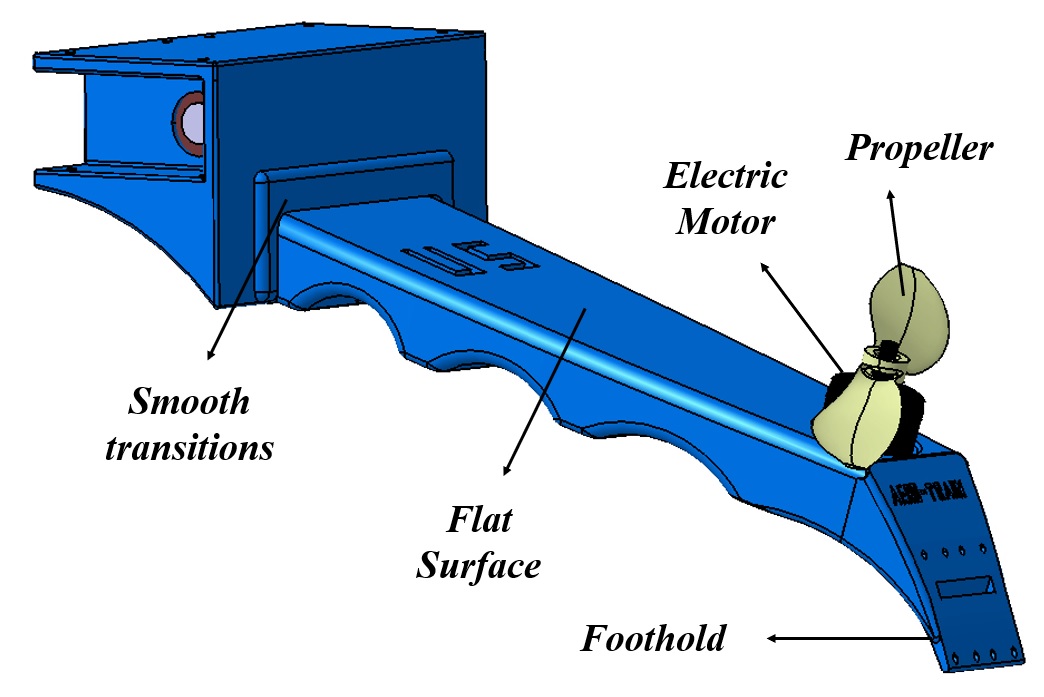}
    \end{minipage}%
    }%
  \end{subfigure}
    \begin{subfigure}[Details of the angles and lengths which the tendons use to bend the arm downwards. Details are provided in \cite{softArm}.]{
    \begin{minipage}[t]{0.96\linewidth}
    \centering
    \includegraphics[width=0.8\textwidth]{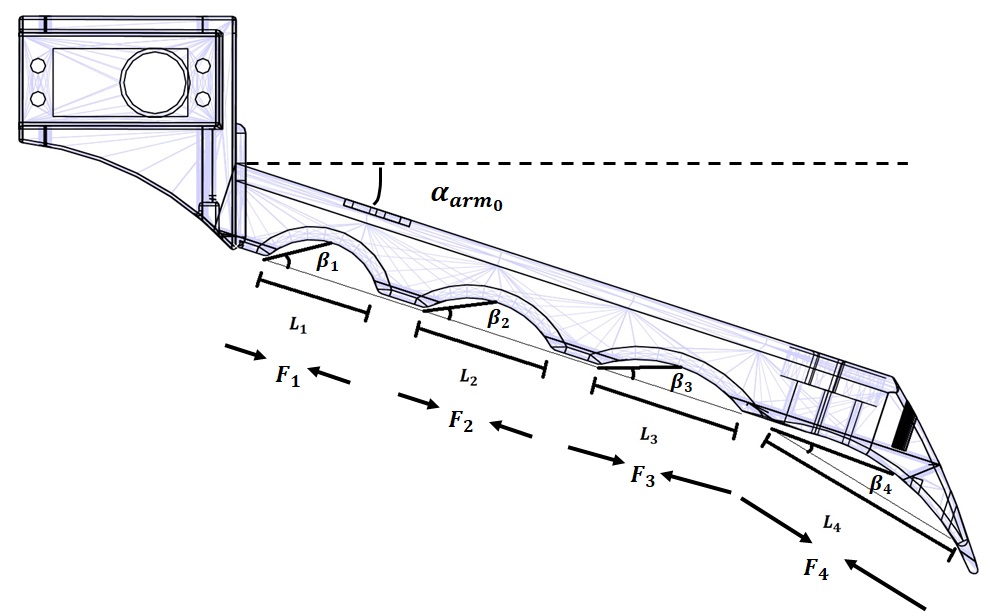}
    \end{minipage}%
    }%
  \end{subfigure}
 \caption{CAD model of the soft arm}
 \label{f:arm}
 
\end{figure}

In the previous work, a detailed mechanical analysis of the material and the most suitable manufacturing techniques for the arm design was carried out \cite{softArm}. In this work, the research is focused on the behavior of the system as a whole, providing insight into the dynamics of the flexible UAV, identifying the different mechanical effects (structural, aerodynamic and aeroelastic) that appear as flexibility increases is fundamental.

Finally, the application to contact inspection in pipelines has been considered given the experience of the authors in previous Oil \& Gas inspection projects (AEROARMS \cite{Aeroarms} and HYFLIERS). For this goal, the previously discussed issue of the vehicle's control is coupled with its ability to deform and attach to the pipeline, as they present opposite requirements. This work intends to understand whether a high enough degree of flexibility can be obtained so that the pressure force by contact with the pipe allows the UAV to stabilize, while assuring the flyability.



\subsection{Soft aerial robot concept} 

This work proposes the development of an innovative aerial robotic vehicle. This disruptive design is just the first step towards the use of flexible materials and soft actuators in UAVs. The aerial robot (Figure \ref{f:UAV}) is a modular and flexible vehicle designed to solve the problems described in Section II-A. The
three main requirements that should be addressed by the
aerial robot are:

1) The entire structure of the aerial robot must be 3D-printed in flexible filament (TPU 70A) which is defined as soft by the ASTM D2240-00 standard.

2) Despite being flexible, the UAV must be able to be stable enough to carry out inspection tasks from the air, without vibrations that compromise the results.

3) The UAV should be able to accomplish a safe land over a pipe and stabilise over it without the need for any auxiliary clamping systems, only using his own deformable arms.




\begin{figure}[!tbp]
  \begin{subfigure}[Configuration A: Rotors 1\&4 rotate CW, Rotors 2\&3 rotate CCW]{
    \begin{minipage}[t]{0.96\linewidth}
    \centering
    \includegraphics[width=\textwidth,scale=1]{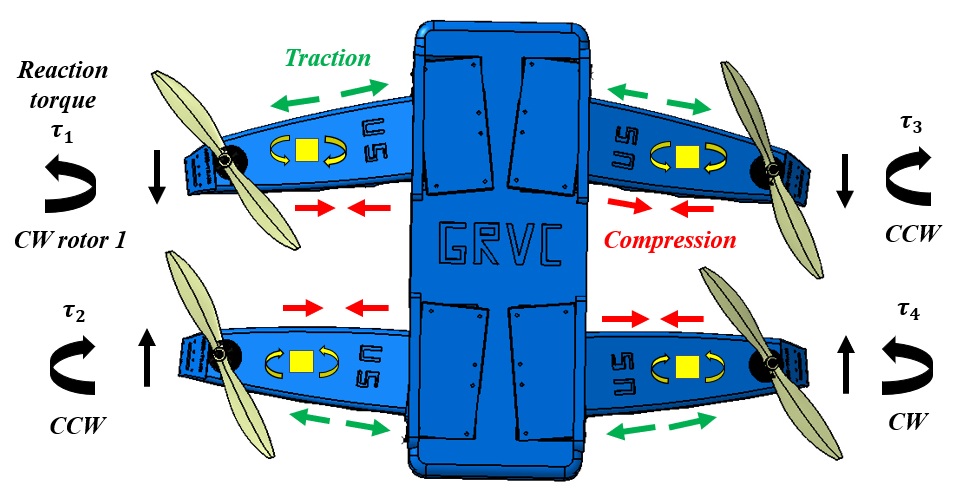}
    \end{minipage}%
    }%
  \end{subfigure}
    \begin{subfigure}[Configuration B: Rotors 1\&4 rotate CCW, Rotors 2\&3 rotate CW]{
    \begin{minipage}[t]{0.96\linewidth}
    \centering
    \includegraphics[width=\textwidth,scale=1]{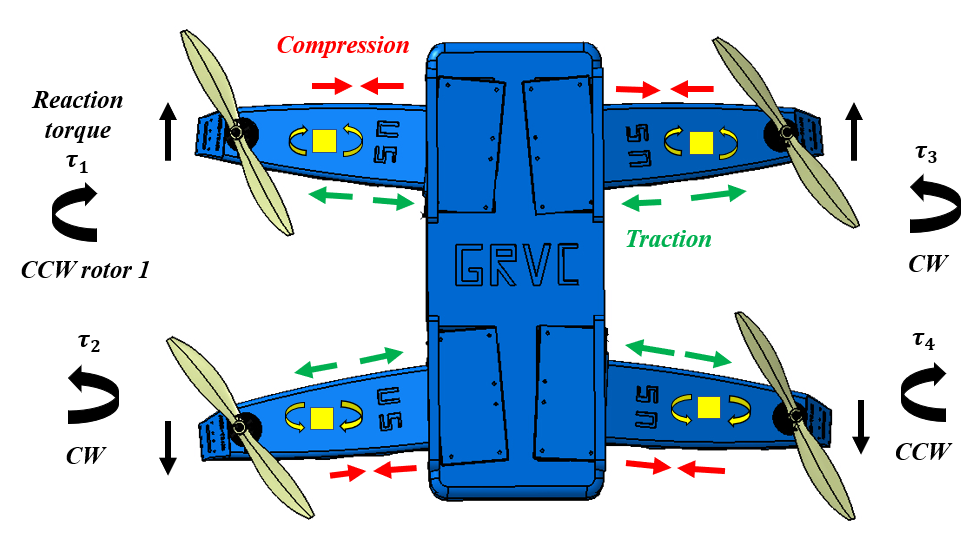}
    \end{minipage}%
    }%
  \end{subfigure}
 \caption{CAD model of the soft UAV. Analysis of the deformations in the horizontal plane in stationary state as a function of the distribution of rotation senses of the rotors}
 \label{f:def}
 \vspace{-5mm}
\end{figure}

\section{Mechanical design}\label{sec:mechanical}


In this section, the mechanical design of the UAV is described. For this purpose, the results from the previous work \cite{softArm} are used, in which a mechanical study of an isolated soft propelled arm was performed. In the present work, the aim focuses on designing a flexible platform with great impact energy absorption, while being compact enough to avoid vibrations and dampen undesired mechanical effects in flight.


\subsection{Soft arm overview}


For the design of the arm, theoretical analyses, mechanical simulations (structural and computational fluid dynamics), and experiments on a test bench have been carried out in \cite{softArm}. However, this article focuses on the behavior and control of the system as a whole.


The CAD design of the arm is shown in Figure \ref{f:arm} along with the main components. The propulsive equipment is composed of an electric motor (DJI
2312E), providing 450g of nominal thrust using a 4S (14.8V) battery, an electronic speed controller (DJI 430 Lite ESC) and a 10-inch DJI plastic propeller to make it safer than carbon fiber. The arm is also equipped with soft tendons on its underside, responsible for generating compression forces that cause the arm to bend downwards. These tendons are composed of nylon threads that are wound on a 3D-printed reel actuated by a HITEC MG996R servomotor which provides a maximum torque of 35 kgcm.


Finally, the arm is equipped with an FSR contact force sensor from the manufacturer Interlink Electronics, which allows analyzing the level of attachment to the pipe. On the other hand, an inertial measurement unit (IMU BNo055) has been used to determine the deflection angle of the arm $\alpha_{arm}$.

\begin{figure}[t]
\includegraphics[width=0.47\textwidth,scale=0.25]{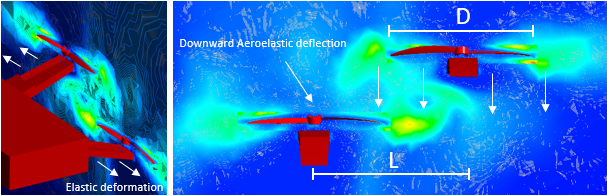}
\caption{Illustration of the aeroelastic effects experienced by the arms of the UAV. Typical design criteria for rigid UAVs establishes that $L/D \geq 1.1$, while in the flexible case $L/D \geq 1.4$ at least, to avoid inefficiencies}
\label{aero}
\end{figure}

Mechanical studies developed in \cite{softArm} led to the following polynomial model for the arm deflections as a function of thrust (T) and flexibility ($\rho_{TPU}$), which constants A1, A2, B1, and B2 were determined from experimental results. In this work, the power loss of the battery is also considered, following a behavior determined by the experimental constant $C_P$ that varies between 1.6 and 1.8:
\begin{equation}
\label{model_arm}
\alpha = \alpha_0+((A_1+\rho_{TPU}*A_2)*T+(B_1+\rho_{TPU}*B_2)T^2)(\frac{P}{P_0})^{C_P}
\end{equation}

\begin{table}[b]
\centering
\begin{tabular}{c c c c}
\hline
A1 & A2 & B1 & B2 \\
2.4387 & -0.1997 & -0.162 & 0.0151 \\
\hline
\end{tabular}
\caption{Arm deflection coefficients for Equation 1}
\vspace{-5mm}
\end{table}

\subsection{UAV design}

CAD design of the UAV is shown in Figure \ref{f:def}. The arms are integrated into the main platform through simple joints, giving rise to a modular and easy-to-assemble design. The main platform has an internal density higher than the arms, and embeds all the necessary electronic equipment and actuators, as well as the autopilot.

The design focuses on minimizing the mechanical effects associated with flexibility. These are both structural and aerodynamic, since there is fluid-structure interaction, known as aeroelastic effects, which will be studied in future work. 

From the structural point of view, the UAV experiences bending loads with respect to its main axis during the roll maneuver, and with respect to its secondary axis during the pitch maneuver. These effects have been dampened by local increases of the infill rate in the bending zones, playing the role of stiffening elements, while being flexible. The critical point of the elastic structure are the torsional loads during the yaw maneuver, which must be carried out as smoothly as possible. Section IV-A deals with stability and control during the yaw maneuver.

In addition to these transient loads while maneuvering, the UAV also experiences continuous loads in stationary flight. These loads appear in the horizontal plane due to the torque generated by the rotors. These bendings occur either $"inwards"$ of the UAV (Figure 4, case A) or $"outwards"$ (Figure 4, case B), depending on the direction of rotation of the rotors. For the present case, the second configuration is more suitable since it leads to less aerodynamic interactions between the flows of the rotors.

These aerodynamic interactions occur fundamentally during the pitch maneuver, in which the difference in inclination between the pair of arms is greater (see Figure \ref{aero}); and eventually it can also occur during yaw maneuvers. The fluid flow from the rotor of the upper arm interacts with the flow of the lower arm and decreases its thrust. In addition, the lower arm undergoes a downward elastic deformation due to the impact of said air flow. This aeroelastic phenomena can potentially be non-stationary, since vibrations are produced in the lower arm.

\begin{figure}[!tbp]
  \begin{subfigure}[Configuration A: Rotors 1\&4 rotate CW, Rotors 2\&3 rotate CCW]{
    \begin{minipage}[t]{0.96\linewidth}
    \centering
    \includegraphics[width=\textwidth,scale=1]{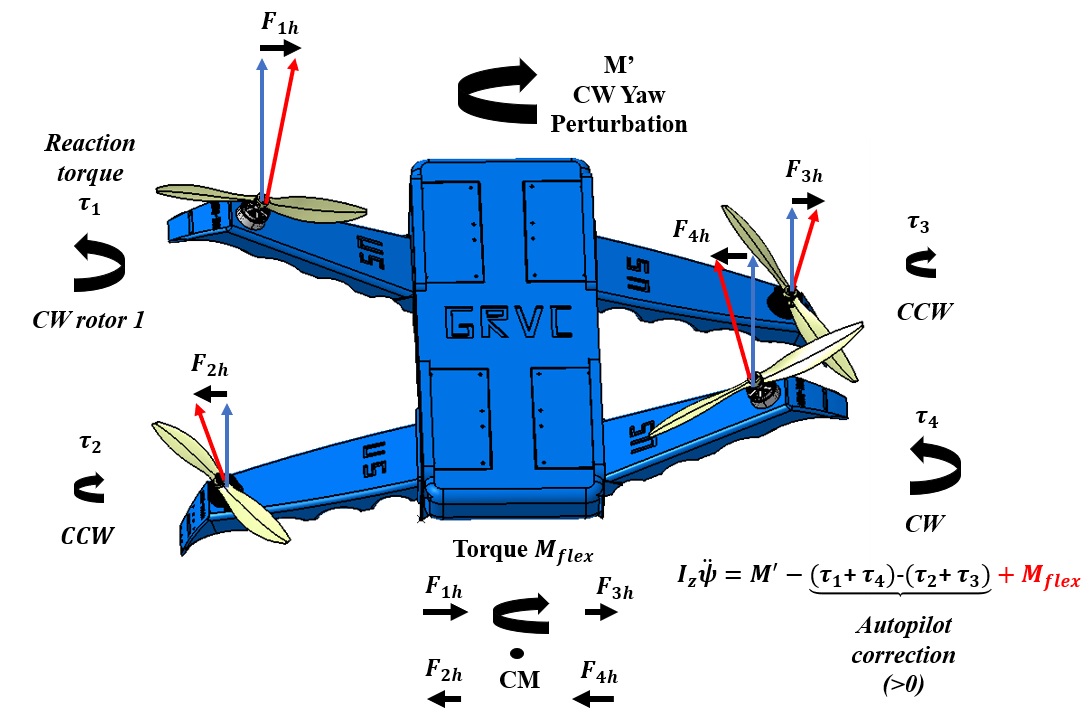}
    \end{minipage}%
    }%
  \end{subfigure}
    \begin{subfigure}[Configuration B: Rotors 1\&4 rotate CCW, Rotors 2\&3 rotate CW]{
    \begin{minipage}[t]{0.96\linewidth}
    \centering
    \includegraphics[width=\textwidth,scale=1]{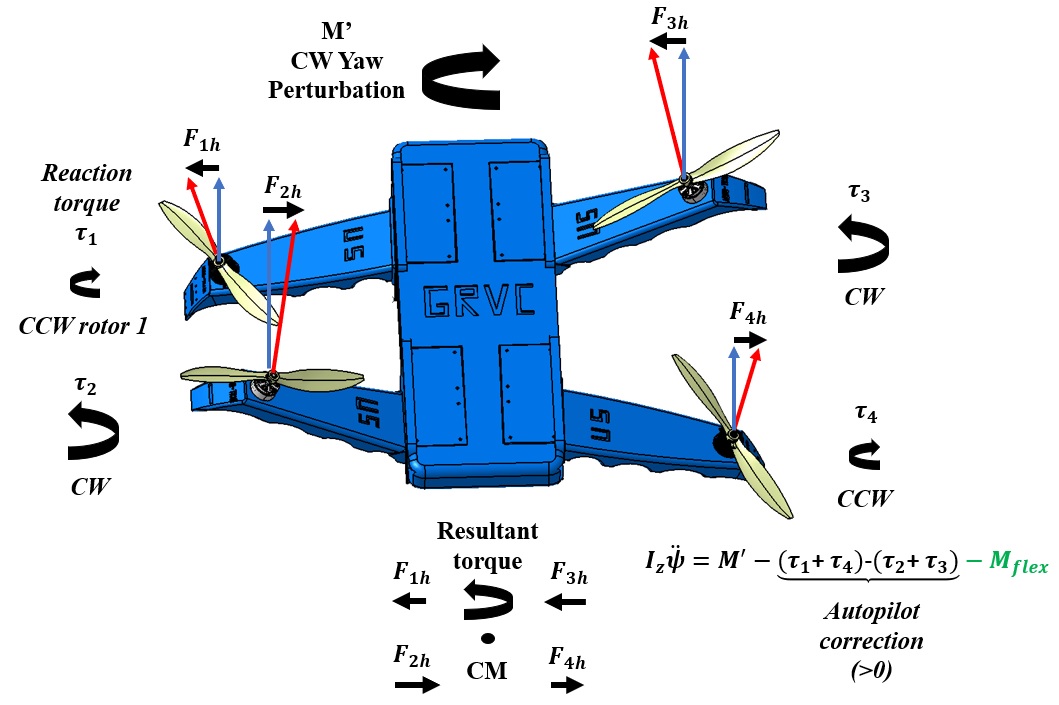}
    \end{minipage}%
    }%
  \end{subfigure}
 \caption{Stability analysis in response to a yaw perturbation $M'_\psi$. Configuration A is unstable since $C_{M_{\tau_{\psi}}}>0$ while Configuration B is stable since $C_{M_{\tau_{\psi}}}<0$}
 \label{f:stab}
 \vspace{-5mm}
\end{figure}



The weight distribution of the UAV is shown in Table II. The total weight of the UAV is around 1.8kg, which is quite encouraging compared to robots used in the literature for pipeline inspection tasks.

\begin{table}[t]
\centering
\begin{tabular}{c c c}
\hline
Equipment & Components & Weight \\
\hline
Arms & 3D-printed & 140g (x4) \\
Platform & 3D-printed & 270g \\
Propulsion & Motor, ESC, propeller  & 100g (x4) \\
Other Mechanisms & Servo, Reel, Bearing  & 70g (x4) \\
Battery & 4S & 185 g \\
Electronics & CUAV V5, RX, PM & 110 g \\
\hline
\end{tabular}
\caption{Weight distribution and components of the flexible UAV }
 \vspace{-6mm}
\end{table}

\section{Control of the system} \label{sec:control}

This aerial system requires specific control strategies for each operation mode (flight and landing controllers, respectively), which are considered independent in this work. Regarding flight operation, the main goal is to provide insight into the \textit{mechanical} stability and dynamics of the UAV while adjusting a commercial autopilot to perform a controlled flight in these conditions. For the landing mode, a PID controller has been used to ensure sufficient clamping to the pipe. 



\subsection{Rotational stability derivatives}

Stability derivatives are measures of how particular forces and moments on a vehicle change when there is a small deflection in the control surfaces such as the ailerons, elevator, and rudder. In the case of UAVs, these control variables are typically the rotational torques $\tau_\phi, \tau_\psi$, and $\tau_\theta$. 

In this work, the flexibility of the vehicle implies forces and moments related to the elastic bending of the arms. Specifically, these arm deflections lead to lateral forces which can potentially generate moments if they are not compensated (see Figure \ref{f:stab}). In this way, while mechanical stability is guaranteed for rigid UAVs, in this work the stability condition implies that the following stability derivatives ($C_{M_{\tau_{\psi}}}$,$C_{M_{\tau_{\theta}}}$, and $C_{M_{\tau_{\phi}}}$) should be negative.

\begin{equation}
\label{psi}
I_z \ddot{\psi}=M'_\psi - \tau_\psi + C_{M_{\tau_{\psi}}} \tau_\psi
\end{equation}
\begin{equation}
\label{theta}
I_y \ddot{\theta}=M'_\theta - \tau_\theta + C_{M_{\tau_{\theta}}} \tau_\theta
\end{equation}
\begin{equation}
\label{phi}
I_z \ddot{\phi}=M'_\phi - \tau_\phi + C_{M_{\tau_{\phi}}} \tau_\phi
\end{equation}

These coefficients $C_{M_{\tau_{i}}}$ determine the moments $M_{flex}$ induced by the flexibility of the UAV when applying a particular $\tau_{i}$, and therefore depend on $\rho_{TPU}$. This condition is naturally fulfilled for pitch and roll ($C_{M_{\tau_{\theta}}}$ and $C_{M_{\tau_{\phi}}}$), as arm deflections contribute to the rotation of the vehicle. However, determining the sign of $C_{M_{\tau_{\psi}}}$ is not trivial, it actually depends on the UAV configuration. As shown in Figure \ref{f:stab}, there is only one possible stable distribution of rotors rotation senses and positions that leads to a negative stability derivative.


\begin{figure}
\includegraphics[width=0.47\textwidth,scale=0.25]{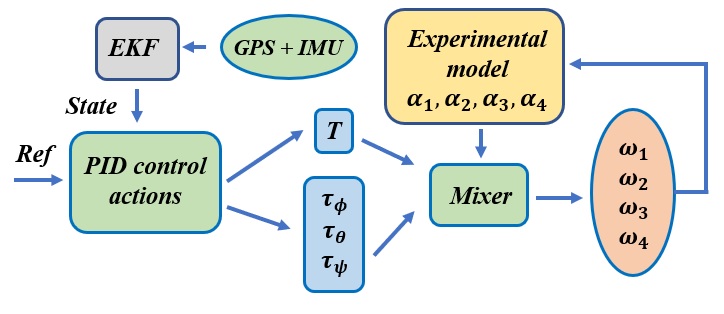}
\caption{Scheme of flight controller operation through an experimental model (Equation 1) for arm deflections that updates the mixer matrix at each timestep}
\label{flightcontroller}
\vspace{-5mm}
\end{figure}

\subsection{Flight controller}


The flight controller is based on the open-source control framework Ardupilot \cite{Ardupilot} following
the standards of an H-quadrotor layout, adapted to a flexible configuration by incorporating the tilt angle of the arms into the mixer through a static experimental model (see Figure \ref{flightcontroller}) and modifying the rotation senses of the rotors to make the system stable (negative stability derivatives). During the flight phase, the servomotors used to bend the arms are deactivated, and manual control is performed through thrust T and torques $\tau_\phi, \tau_\psi$, and $\tau_\theta$, corresponding to the traditional 4 degrees of freedom of the UAV. 

However, the flexibility of the UAV requires accounting for arm deflections $\alpha_i$, as it is shown in Figure 8, since rotors are not coplanar and mixer matrix changes at each timestep. In this way, $\alpha_i$ is considered positive when the arm is up ($T>50\%$ if properly designed) and negative when it is down ($T<50\%$). 

The modified mixer matrix can be seen below. Analyzing the implications of these changes in the translational equations is essential to understand the dynamics of the UAV, and how the thrust of the rotors is projected on the XY plane through a component $f_x$. 


\begin{equation*} 
\begin{bmatrix} 
f_x \\
f_y \\
f_z \\
\tau_\phi \\
\tau_\theta \\
\tau_\psi 
\end{bmatrix}
=
\begin{bmatrix} 
-s{\alpha_1}C_T & -s{\alpha_2}C_T & s{\alpha_3}C_T & s{\alpha_4}C_T  \\
0 & 0 & 0 & 0 \\
c{\alpha_1}C_T & c{\alpha_2}C_T & c{\alpha_3}C_T & c{\alpha_4}C_T \\
dC_T & dC_T & -dC_T & -dC_T \\
dC_T & -dC_T & dC_T & -dC_T \\
-C_Q & C_Q & -C_Q & C_Q
\end{bmatrix}
\begin{bmatrix}
\omega_1^2 \\
\omega_2^2 \\
\omega_3^2 \\
\omega_4^2
\end{bmatrix}
\end{equation*}

For $\alpha_i < 40 ^o$ angles, which occurs for $\rho_{TPU} > 5\%$, arm deflections can be considered quasi-static and obtained in real time from Equation \ref{model_arm} as explained in Figure \ref{flightcontroller}. Outside this range, strong non-linear dynamics appear, and the static model is not valid anymore.  For lower densities causing such large deflections, it would be more suitable to incorporate reinforcement learning techniques to actuate the system. The implementation in the Ardupilot environment (AP-MotorsMatrix.cpp file) is performed calculating $\alpha_i(t+1)$ from $\omega(i)(t)$ (previous timestep) using the experimental model from Equation 1, as it is detailed in Algorithm 1.

The quasi-static arm deflections assumption is considered valid since the dynamics of the arm are slow with regards to changes in $\omega(i)$. Furthermore, arm vibrations are considered negligible due to the dampening properties of TPU at these infill rates. For small $\alpha_i$ angles, the forces $f_x$ and $f_z$ can be written as


\renewcommand{\algorithmicrequire}{\textbf{Input:}}
\renewcommand{\algorithmicensure}{\textbf{Output:}}

\begin{algorithm}[t]
  \caption{Tilt angles calculation and generation of the case in the Ardupilot environment (AP-MotorsMatrix.cpp)}
  \begin{algorithmic}[1]
 
    \Require{Rotational speeds and available power at the previous time step ($w_i(t-1)$, $P(t-1)$), arm deflection coefficients $(A_i)$, infill rate ($\rho_{TPU}$) and initial deflection angles ($\alpha_{i0}$)} 
    \Ensure{Tilt angles $\alpha_i(t)$}                                      

    \For{$k;\ k++;\ |Tf|$}
        \State $T_i(k-1)= C_T \omega_i(k-1)^2$
        \State $\alpha_i(k) = \alpha_{i0} + ((A_1+\rho_{TPU}*A_2)*T(k-1) + (B_1+\rho_{TPU}*B_2)*T(k-1)^2)*(\frac{P(k-1)}{P_0})^{C_P}$
        \If{$\alpha_i(k)>40º$}
            \State $\alpha_i(k)=0º$  \textit{Non-linear region, invalid model}
        \EndIf
    \EndFor
    \State case SoftMultirotor: 
    \For{$i=1:2$}
        \State $add_{motor}(i, \alpha_i(k), CW, 0)$;
    \EndFor  
    \For{$i=3:4$}
        \State $add_{motor}(i, \alpha_i(k), CCW, 180)$;
    \EndFor

    \State{\textbf{return} APMotorsMatrix}
  \end{algorithmic}
\end{algorithm}


\begin{equation}
\label{hypotesisZ}
\begin{split}
\mathbf{f_z} = c{\alpha_1}C_T\omega_1^2 + c{\alpha_2}C_T\omega_2^2
+ c{\alpha_3}C_T\omega_3^2 + c{\alpha_4}C_T\omega_4^2  \mathbf{\approx T}
\end{split}
\end{equation}
\begin{equation}
\label{hypotesisX}
\begin{split}
\mathbf{f_x} = -s{\alpha_1}C_T\omega_1^2 - s{\alpha_2}C_T\omega_2^2 + s{\alpha_3}C_T\omega_3^2 + s{\alpha_4}C_T\omega_4^2 \\
\mathbf{\approx -\alpha_1 T_1 - \alpha_2 T_2 + \alpha_3 T_3 + \alpha_4 T_4}
\end{split}
\end{equation}

Therefore, the governing equations for the translational dynamics (neglecting aerodynamic forces) of the UAV are

\begin{equation}
\begin{split}
\label{tras_dynamics_x}
m\Dot{v}_x = - (s\psi s\theta c\phi - c\psi s\phi)T + \\
(c\theta s\psi)(-\alpha_1 T_1 - \alpha_2 T_2 + \alpha_3 T_3 + \alpha_4 T_4) 
\end{split}
\end{equation}
\begin{equation}
\begin{split}
\label{tras_dynamics_y}
m\Dot{v}_y = - (c\psi s\theta c\phi + s\psi s\phi)T + \\ 
(c\theta c\psi)(-\alpha_1 T_1 - \alpha_2 T_2 + \alpha_3 T_3 + \alpha_4 T_4)
\end{split}
\end{equation}
\begin{equation}
\begin{split}
\label{tras_dynamics_z}
m\Dot{v}_z = - (c\psi s\theta c\phi + s\psi s\phi)T - \\                    
(s\theta)(-\alpha_1 T_1 - \alpha_2 T_2 + \alpha_3 T_3 + \alpha_4 T_4)
\end{split}
\end{equation}

\begin{figure}
\includegraphics[width=0.47\textwidth,scale=0.25]{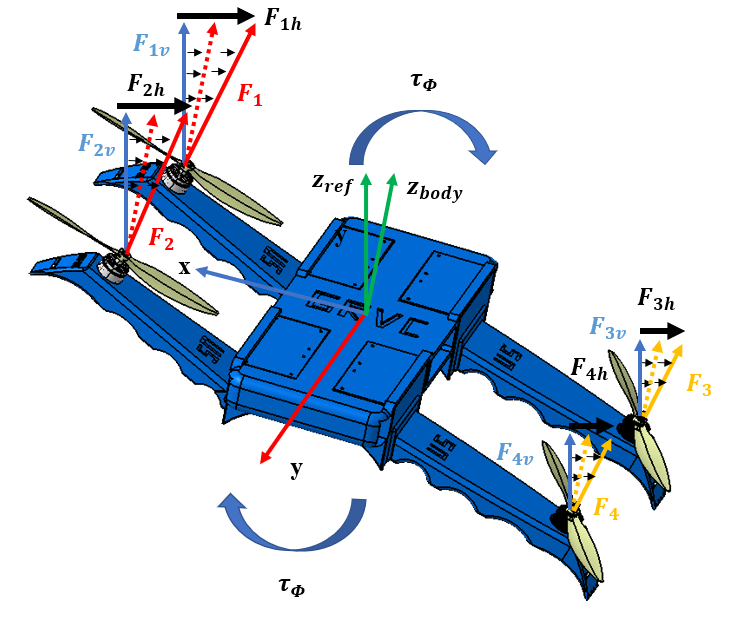}
\caption{Physical interpretation of the flexibility effect on lateral dynamics. The horizontal component of the effective thrust (red solid arrows) is greater than in the rigid case (red dashed arrows) since the effective angle ($\vartriangle\alpha+\phi$) is bigger.}
\label{lat}
\end{figure}

To physically interpret these equations, consider a roll motion of the UAV while it points north ($\psi=0$) in response to a torque $\tau_\phi$. Consider also that the pitch is ($\theta=0$) during this maneuver. For this particular case $F_1 = F_2 > F_3 = F_4$, as the autopilot generates differential thrust to achieve the desired $\tau_\phi$, and, therefore, $\alpha_1 = \alpha_2 > \alpha_3 = \alpha_4$. Incremental variables $\vartriangle\alpha$ and $\vartriangle T$ are then defined as $\vartriangle\alpha=\alpha_1-\alpha_3$ and $\vartriangle T=T_1-T_3$. Introducing this into equation \ref{tras_dynamics_x}

\begin{equation}
\label{tras_dynamics_y_simple}
m\Dot{v}_x = \underbrace{s\phi T}_{Rigid} +  \underbrace{\vartriangle\alpha \vartriangle T }_{Flexible}
\end{equation}

which shows that the flexibility of the UAV entails an extra lateral force during the roll maneuver. The importance of this force will depend on the degree of flexibility, for a rigid vehicle $\vartriangle\alpha = 0$ and the traditional equation $m\Dot{v}_x = s\phi T$ is recovered. These concepts can be graphically interpreted in Figure \ref{lat} and are extensible to a lateral motion in the y-direction when a pitch torque $\tau_\theta$ is applied.



\subsection{Landing on pipelines}

\begin{figure}
\includegraphics[width=0.47\textwidth,scale=0.25]{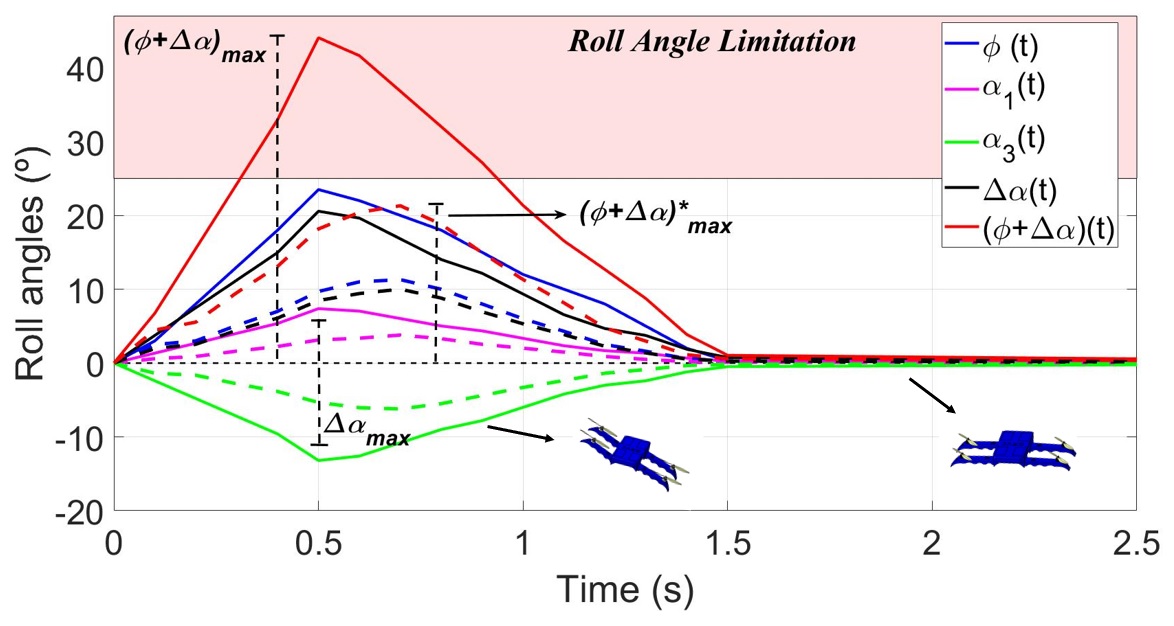}
\caption{Flight experiment for $\rho_{TPU}=8\%$. Evolution of $\phi(t)$ (blue lines) and $\alpha_i(t)$ (magenta and green lines) during the roll maneuver with (dashed lines) and without (solid lines) autopilot corrections. Evolution of the effective angle $\phi(t)+\vartriangle\alpha(t)$ (red lines) for both cases.}
\label{Roll}
\end{figure}

The landing maneuver is initiated by the operator after touching the surface (otherwise the vehicle would destabilize) with the air propulsion system still on and a throttle of around 30 $\%$ to provide extra stability during the operation. The closing process is a closed-loop control phase using the contact force FSR sensors and tendons actuated by servomotors in order to guarantee an adequate attachment on pipes of different diameters. The target contact pressure is $950 N/m2$ and the maximum allowable servomotor force is $F_{max}=25 kgcm$, which have been carefully selected so that the UAV is stabilized over the pipeline without risking a possible rupture of the tendons. Note also that the stability not only comes from the contact of the arms, but also from lowering the center of gravity of the vehicle closer to the axis of the pipe, unlike previous designs \cite{MHYRO,PRamon2019}.

The maneuver can be performed with both two or four arms. The first case is more interesting since it guarantees attachment while leaving the other two arms free for inspection tasks. Note that the aerodynamic ground effect (on a curved surface) is present during this maneuver, which tends to detach the arms from the pipe. For this reason, if only two arms are used, flow interactions after bouncing in the pipe are reduced. In addition, take-off is less complex as there are two arms that provide vertical thrust. 

The PID controller has been adjusted experimentally, until the desired response has been obtained, focusing on avoiding contact pressure peaks. The proportional constant is adjusted depending on the type of surface. Then, an integral action has been added to reduce the error in permanent regime, since FSR sensor measurements have a significant error ($\approx$50 N/m2). Finally, the derivative constant has been selected to dampen overshoots and oscillations. 




\section{Results}

This section is intended to validate the flying capabilities of the UAV for different flexibilities and the ability to perch on pipelines. Discussions on vehicle performance and optimum degree of flexibility are provided. Finally, the main features of the UAV are compared with conventional rigid aerial vehicles.



\subsection{Flight operation overview}

To analyze the lateral dynamics of the UAV, a new parameter $\kappa$ (Equation 11) is defined, which measures the importance of the flexible term in equation \ref{tras_dynamics_y_simple} with respect to the rigid term. The influence of flexibility manifests itself in arm deflections $\Delta\alpha_{max}$, as shown in Figure \ref{Roll} for the case $\rho_{TPU}=8\%$. These deflections generate an extra lateral force since the total effective angle is bigger than expected $(\phi+\Delta\alpha)_{max}$.

\begin{equation}
\label{kappa}
\kappa = \frac{\vartriangle\alpha \vartriangle T}{\sin{\phi}T}
\end{equation}

Flight experiments were performed with $\phi_{max}=25^o$, which corresponds to $\frac{\vartriangle T}{T}\approx\frac{3}{5}$. Autopilot corrections allow to reduce the aggressiveness of the roll maneuver, reducing the total effective angle to $(\phi+\Delta\alpha)*_{max}$. These flight experiments have been repeated for other $\rho_{TPU} (\%)$ values, which are shown in Table III. $\kappa$ can be interpreted as the influence on the lateral force and $\frac{(\phi+\Delta\alpha)_{max}}{(\phi+\Delta\alpha)*_{max}}$ as the direct influence on the roll angle.

\begin{table}[b]
\centering
\begin{tabular}{c c c c}
\hline
$\rho_{TPU} (\%)$ & $\kappa$ & $\Delta\alpha_{max} (^o)$ & $\frac{(\phi+\Delta\alpha)_{max}}{(\phi+\Delta\alpha)*_{max}}$ \\
\hline
6 & 0.714 & 26.6 & 2.59\\
8 & 0.569 & 21.3 & 2.08 \\
10 & 0.488 & 16.8 & 1.63\\
\hline
\end{tabular}
\caption{Influence of the infill rate $\rho_{TPU}$ on the lateral force induced by the flexibility of the arms}
\vspace{-5mm}
\end{table}

\subsection{Inspection operation analysis}

\begin{figure}
\includegraphics[width=0.47\textwidth,scale=0.25]{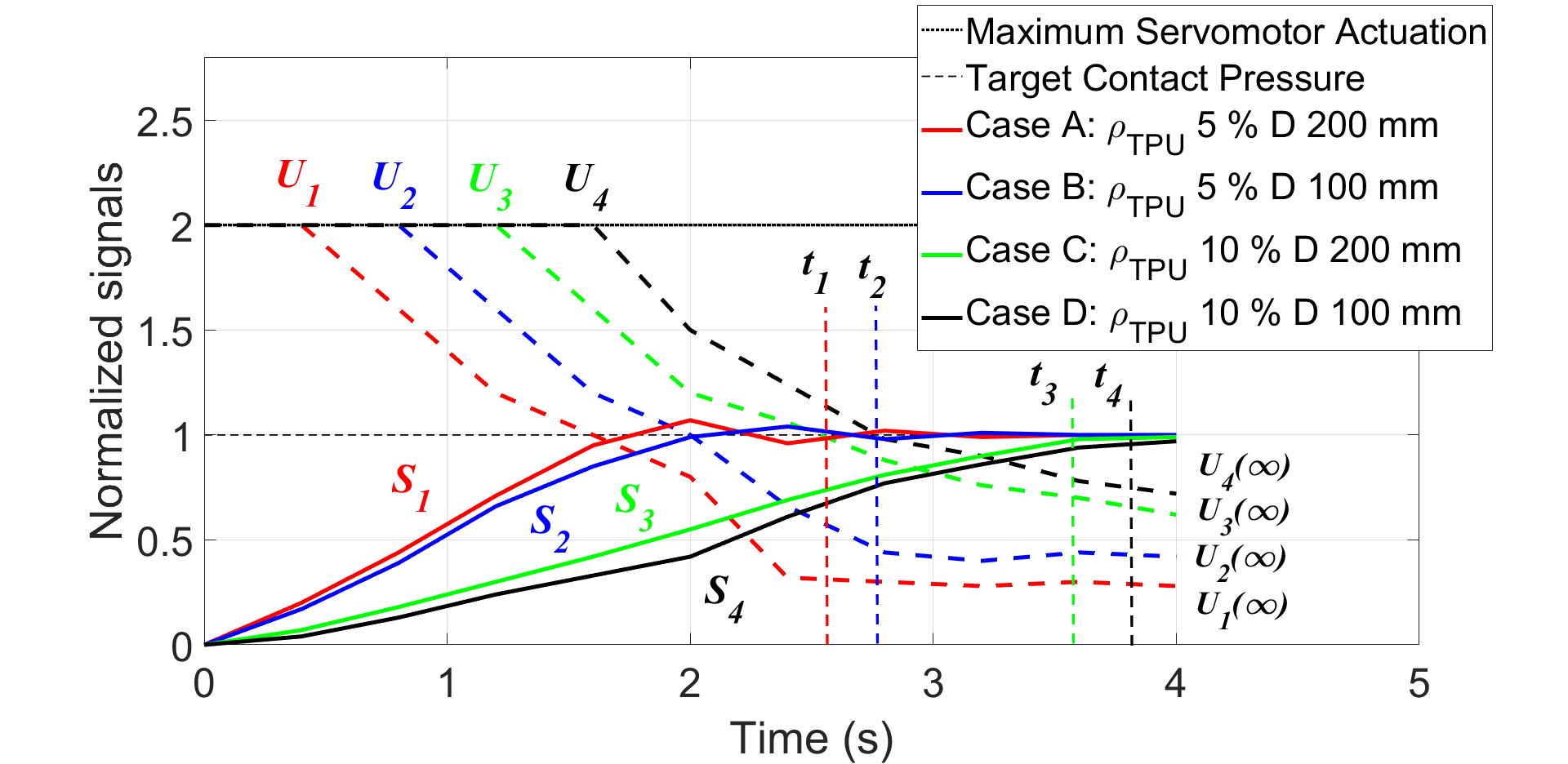}
\caption{State (normalized contact pressure $s(t)=\frac{P}{P_{max}}(t)$, solid lines) and actuation (servomotor force $u(t)=\frac{F}{0.5F_{max}}(t)$, dashed lines) signals in the PID controller for the servomotors, for different pipe diameters D and internal densities $\rho_{TPU}$.}
\label{PID}
\end{figure}

The closing mechanism has been studied on a test bench for different pipe diameters (D) and material flexibilities ($\rho_{TPU}$), measuring both closure times $t_i$ and servomotor force requirement in stationary state $U_i(\infty)$, which is a measure of the energy consumption during the inspection operation. Closure times (see Figure \ref{PID}) are bigger for small pipe diameters and high internal densities. The same conclusions can be drawn for $U_i(\infty)$. Therefore, inspection times will be shorter the less resistance the material opposes, for the same battery capacity. The numerical values obtained are shown in Table IV.

\begin{table}[H]
\centering
\begin{tabular}{c c c c}
\hline
$t_1$ & $t_2$ & $t_3$ & $t_4$ \\
2.61s & 2.83s & 3.65s & 3.85s \\
\hline
$U_1(\infty)$ & $U_2(\infty)$ & $U_3(\infty)$ & $U_4(\infty)$ \\
9.2kgcm & 8.24kgcm & 5.19kgcm & 3.93kgcm \\
\hline
\end{tabular}
\caption{Closure times of the arm $t_i$ and force required by the servomotor $U_i(\infty)$ to maintain contact at the steady state}
\vspace{-5mm}
\end{table}

\subsection{Material degradation}

Material degradation due to the fatigue has been found after 50-100 bending cycles, which can potentially alter the vehicle's stability. Fiber ruptures lead to arm deflections above expectations for the same level of thrust (Equation 1 is not valid anymore).


In flight, the autopilot tries to compensate this unexpected deflection by raising the neighbour arm to equalize the horizontal forces and the UAV remains stabilized in position. However, this compensation entails non-equal rotational speeds of the rotors, leading to yaw perturbations. This means that the autopilot must be continuously correcting in yaw to be stable. For flexibilities $\rho_{TPU}<6\%$, the autopilot is not able to correct these perturbations in real time.

\subsection{Overall performance comparison}

In this section, the performance of the flexible UAV is analyzed, throwing a comparison with a rigid UAV. For this purpose, the following characteristic times are defined:

\begin{itemize}

\item Maximum flight time (FT) calculated by measuring the average electrical consumption $(C_M)$ of the battery (4S of 1800mAh) in a 1-minute period, for hovering conditions (stationary flight). Flight time can therefore be calculated as the ratio between the battery capacity and the average consumption $C_M$. FT increases with the flexibility of the vehicle due to the lower density and weight. However, at very low densities ($\rho_{TPU}=4\%$), excessive arm deflections decrease the effective thrust and autonomy (see Table V).

\item Maximum inspection time (IT) that the UAV can remain attached to the pipeline performing the inspection task. It is calculated in a similar way, obtaining the consumption of the servomotor to maintain the closing forces $U_i(\infty)$ of table IV, given the battery capacity (2S of 1200mAh).

\end{itemize}

\begin{table}[t]
\centering
\begin{tabular}{c c c c}
\hline
$\rho_{TPU}(\%)$ & Flight Time & Inspection time & Specific strength ($\frac{kNm}{kg}$)\\
\hline
$4$ & $10.7$ min & $16.7$ min & $0.9$ \\
$6$ & $15.8$ min & $7.5$ min & $2.3$ \\
$8$ &  $14.7$ min & $6.4$ min & $4.5$ \\ 
$10$ &  $14.1$ min & $2.5$ min & $10.4$ \\ 
\hline
\end{tabular}
\caption{SOPHIE's flight time, inspection time and specific strength (obtained from \cite{softArm}) for different infill rates $\rho_{TPU}$}. 
\vspace{-8mm}
\end{table}

In addition to these characteristic times, to assess the qualities of the UAV, its flyability and control (through the $\kappa$ parameter) and the safety of the UAV (through the material's specific strength which measures the ability to store energy before fracture) should also be considered.




From these results, it is concluded that optimal infill rates are between 6$\%$ and 8$\%$, where the material is flexible enough to adapt to pipes and hold pressure for a long time (at least 6-7 mins), while it is controllable through the models and corrections proposed in this work. For $\rho_{TPU}<5\%$, the behavior is too complex and non-linear and it is proposed to switch to machine learning techniques and add extra actuation systems. At densities greater than 10$\%$, the energy required to remain attached to the pipeline is excessively large and therefore inspection times are scarce.


\begin{table}[b]
\centering
\begin{tabular}{c c c c}
\hline
Aerial Robot & Flight Time & Inspection time & Specific strength \\
\hline
SOPHIE & $16.1$ min & $7.1$ min & $2.73$ $\frac{kNm}{kg}$ \\
Semi-Rigid UAV & $16.8$ min & $1.7$ min & $43$ $\frac{kNm}{kg}$ \\
MHYRO \cite{MHYRO} & $8.85$ min & $4.75$ min & $220$ $\frac{kNm}{kg}$ \\
Soft Gripper \cite{PRamon2019} &  $10.1$ min & $7.4$ min & $78.5$ $\frac{kNm}{kg}$ \\ 
\hline
\end{tabular}
\caption{Numerical values for flight time, inspection time and specific strength for different aerial robots}
\end{table}

Therefore, to perform a comparison with a rigid UAV, the final prototype has been manufactured with an infill rate of $\rho_{TPU}=6.9\%$. For the rigid UAV, previous prototypes designed by the authors for pipeline inspection have been taken as a reference \cite{PRamon2019, MHYRO}. The weight of the vehicle has been reduced by 38$\%$ and 47$\%$ compared to those traditional rigid quadrotors with an auxiliary system designed to perch on pipelines, leading to a significant increase in flight time. 

Regarding inspection, the use of deformable arms also allows inspection times to be increased by 33 $\%$ compared to \cite{MHYRO}. These results are quite encouraging and justify the use of a flexible UAV despite the consequences on the controllability of the UAV that have been shown in this work.

Furthermore, the tenacity of the soft material is a couple of orders of magnitude higher than typical carbon fiber rigid frames. This guarantees to avoid breakage in case of impact with the ground or the surroundings. In addition, it significantly reduces the risks in the event of a collision with a human being.

In comparison with a semi-rigid drone 3D-printed in both TPU and PLA, the conclusion is that SOPHIE's small efficiency losses (which barely exceed 6$\%$) are compensated by much longer inspection times.




\section{CONCLUSIONS}

In this paper, a novel flexible aerial vehicle concept has been proposed. It has been entirely 3D-printed in flexible material, with a modular and lightweight design. This work has proven the flyability of the vehicle for infill rates as low as 6$\%$.

In this process, elastic effects have been studied, and a specific design has been proposed to dampen undesired bending motions through adjustments in the infill rate distribution, while maintaining the flexibility of the assembly. A strong interaction between the fluid flows and the flexible structure has been found, which will be studied in subsequent works.

Furthermore, it has been highlighted the importance of the choice of the rotation senses of the rotors, and their influence on the mechanical behavior and the stability of the UAV. 


The ability to perform full-body perching has been demonstrated, both with all four arms and with only two, leaving the others free for inspection tasks. The repeatability of the experiments is very high in the two-arm case, while in the four-arm case it depends on the pilot skills. Another potential application of this flexible UAV concept is grasping.

For the control of the UAV, a quasi-static approach has been used for the arm deformation, which is only valid in a certain range of flexibilities. This model has been fed back to the autopilot in real time to guarantee the stability of the vehicle.

The path of the authors' research goes towards the application of AI techniques to improve the control of this type of UAVs at higher flexibilities, where strong non-linear elastic dynamics appear. Reinforcement learning techniques would also be interesting to deal with material degradation. 

Within the Aero-Train project, the goal is to equip the flexible UAV with an ultrasonic sensor to detect signs of corrosion by measuring the thickness of the pipe. This is done in collaboration with the company ENI in Italy.

\addtolength{\textheight}{-12cm}   




\section*{ACKNOWLEDGMENTS}

This work has been developed within the framework of the AERO-TRAIN project, a Marie-Sklodowska-Curie Innovative Training Network (ITN) project  (Grant agreement 953454). The funding of the HYFLIERS  project (H2020-2017-779411) is also acknowledged.  We thank Robotics, Vision and Control Group (GRVC) for supporting us during this work.


\bibliographystyle{ieeetr}
\bibliography{references.bib}

\end{document}